\documentclass[thmsa,12pt]{article}

\usepackage{amssymb}
\usepackage{float}
\usepackage{makeidx}
\usepackage[dvips]{graphicx}
\usepackage{indentfirst}

\title{\bf Measurements of Multicomponent Diffusion Coefficients for Lysozyme
Chloride in Water and Aqueous Na$_2$SO$_4$}

\author{Daniela Buzatu\thanks{Physics Department,
Politehnica University, Bucharest, 77206, Romania.}, Emil
Petrescu\footnotemark[1], Florin D. Buzatu\thanks{Department of
Theoretical Physics, National Institute for Physics and Nuclear
Engineering, Bucharest-M\u{a}gurele, 76900, Romania.} \\
John G. Albright\thanks{Department of Chemistry, Texas Christian
University, USA}}

\date{}

\begin{document}

\maketitle

\setlength{\unitlength}{1mm}
\newcommand{\pl}{\partial}
\newcommand{\ep}{\varepsilon}

\begin{abstract}
\noindent This paper presents a diffusion experimental study for
ternary lysozyme-Na$_2$SO$_4$-water system, from moderate
precipitant concentrations into the supersaturated region and
provides a complete set of four diffusion coefficients. These data
are important in order to provide accurate models of protein
diffusion with applications in growth of protein crystals  for
X-ray diffraction studies. All three-component mutual-diffusion
experiments reported here were performed by Rayleigh
interferometry at pH$=4.5$, T$=25^o$ C and at a mean lysozyme
concentration (average of top and bottom solution concentrations)
of 0.6 mM (8.6 mg/mL). Four experiments, with different
combinations of protein and Na$_2$SO$_4$ concentration
differences, were performed at each of five mean Na$_2$SO$_4$
concentrations (0.1, 0.25, 0.5, 0.65 and 0.8 M), for a total of 20
experiments. In addition, we have measured dynamic
light-scattering diffusion coefficients of the ternary system
lysozyme chloride-Na$_2$SO$_4$-water.

\end{abstract}

\newpage
\noindent {\bf Motivation} \\

\noindent The diffusion of protein is one of the fundamental
processes occurring in biological systems, and it is also an
important step in the crystallization mechanism. Obtaining protein
crystal of good structural quality is often the main issue for
three dimensional atomic resolution structure studies of
biomacromolecules. Crystallization  is an intrinsically
non-equilibrium process, and concentration gradients will occur
around the crystal. The protein crystallizes, reducing its
concentration at the moving face of the growing crystal and
creating a protein gradient between the bulk solution and the
crystal. This gradient in turn causes multicomponent diffusive
transport of protein and precipitant. Diffusion in protein crystal
growth inevitably occurs under conditions for which no species has
an uniform concentration raising the issue of multicomponent
diffusion. \\
\noindent Crystallization experiments conducted under microgravity
conditions have yielded protein crystals that provided diffraction
data of significantly higher resolution than the best crystal of
these proteins grown under normal conditions \cite{1}. The
difference between microgravity and normal gravity is the
magnitude of the buoyancy forces. Commonly, this difference is
assigned to some consequences of the reduction of buoyancy driven
convection in the micrigravity conditions. With convection, the
lysozyme concentration in the bulk solution is more uniform
\cite{2} and the probability for nucleation of parasitic crystals
is strongly reduced. \\
\noindent The complete description of an $n$-solute system
requires an $n\times n$ matrix of diffusion coefficients relating
the flux of each solute component to the gradients of all solute
components \cite{3}. The importance of other species on protein
diffusion follows from the one-dimensional flux relations
\cite{3}:
\begin{eqnarray}
-J=\sum_{j=1}^n(D_{ij})_v\partial C_j/\partial
x\;\;\;\;\;\;\;i=1,....,n
\end{eqnarray}
in which the cross-term diffusion coefficients (off-diagonal
elements $(D_{ij})_v\;i\neq j$) can be positive or negative. In
ternary systems ($n=2$), our case, the one-dimensional flux
relations could be written as:
\begin{eqnarray}
-J_1 & = & (D_{11})_v\frac{\partial C_1}{\partial
x}+(D_{12})_v\frac{\partial
C_2}{\partial x} \\
-J_2 & = & (D_{21})_v\frac{\partial C_1}{\partial
x}+(D_{22})_v\frac{\partial C_2}{\partial x}
\end{eqnarray}
where $J_1$ and $J_2$ - the protein flux and respectively salt
flux, $(D_{11})_v$ and $(D_{22})_v$ - the main-term diffusion
coefficients relating to the flux of component to its own
concentration gradient, and $(D_{12})_v$ and $(D_{21})_v$ - the
cross-term diffusion coefficients relating the flux of each
component to the gradient of the other. For some systems \cite{4,
5}, a cross-terms $(D_{ij})_v$ can have considerably larger
magnitude than the main-terms $(D_{ii})_v$, as our measurements
show for the Lys-Na$_2$SO$_4$-Water system described here and in
agreement with the results about Lys-NaCl-Water system \cite{6}.
The index $v$ from the diffusion coefficients shows that the
experiment were done under the assumption the volume change on
mixing and changes in concentrations across the diffusion boundary
were small. Consequently, with a good approximation, the measured
diffusion coefficients may be considered to be for the
volume-fixed reference frame \cite{7} defined by:
\begin{eqnarray}
\sum_{i=0}^nJ_i\bar{V}_i=0
\end{eqnarray}
where $\bar{V}_i$ is the partial molar volume of the $i$th
species, and the subscript 0 denotes the solvent. \\
The importance of multicomponent diffusion has been recognized in
the crystal growth community \cite{8,9} and a crystal growth model
has properly accounted for multicomponent diffusive transport in
lysozyme chloride-NaCl-water system \cite{6, 10}. The experimental
multicomponent diffusion coefficients are essential for accurate
modeling of protein transport, especially in view of the very
large cross-term coefficient $(D_{21})_v$ reported here. Moreover,
the concentration of supporting electrolyte dependence of all the
diffusion coefficients should be important for supersaturation
region and also for its directly contribution to the protein flux.
\\ \\

\noindent {\bf Experimental section} \\

\noindent All the experimental work was performed al Texas
Christian University, in the Chemistry Department. \\ \\
\noindent {\bf Materials}. All the materials, solution preparation
procedures, apparatus and density measurement procedures are
described in the work \cite{6}. We used a hen egg-white lysozyme,
recrystallized six times purchased from Seikagaku America.  \\
The molecular mass of the lysozyme solute, $M_1$, was taken as
14307 g/mol, and this value \cite{11} was used to calculate all
concentrations after correction for the moisture and chloride
content. Buoyancy corrections were made with the commonly used
lysozyme crystal density \cite{12,13,14} of 1.305 g/cm$^3$.\\
The molecular mass of water, $M_o$, was taken as 18.015 g/cm$^3$
and the molecular mass of Na$_2$SO$_4$, M$_2$, was taken as
142.037 g/mol. \\
\noindent Mallinckrodt reagent HCl ($\sim$ 12 M) was diluted by
half with pure water and distilled at the constant boiling
composition. This resulting HCl solution ($\sim $ 6 M) was then
diluted (pH 1.2) and used to adjust the
pH of solution.\\

\noindent {\bf Preparation of Solutions}. All solutions were
prepared by mass with appropriate buoyancy corrections. All
weighings were performed with a Mettler Toledo AT400
electrobalance. Since the as-received lysozyme powder was very
hygroscopic, all manipulations in which water absorption might be
critical were performed in a dry glove box. Stock solutions of
lysozyme were made by adding as-received protein to a pre-weighted
bottle that had contained dry box air, capping the bottle, and
reweighing to get the weight and thus mass of lysozyme. Water was
added to dissolve the lysozyme, and the solution was weighed. An
accurate density measurement was made and used to obtain the
molarity of the stock solution. \\
\noindent For ternary experiments, precise masses of Na$_2$SO$_4$
were added to flasks containing previously weighed quantities of
lysozyme stock solutions. These solutions were mixed and diluted
to within 10 cm$^3$ of the final volume. The pH was adjusted, and
the solutions were diluted to their final mass. \\ \\
{\bf Measurements of pH}. The pH measurements were made using a
Corning model 130 pH meter with an Orion model 8102 combination
ROSS pH electrode. The meter was calibrated with standard pH 7 and
pH 4 buffers and checked against a pH 5 standard buffer.
\\ \\
{\bf Density Measurements}. All density measurements were made
with a Mettler-Paar DMA40 density meter, with an RS-232 output to
a Apple $\Pi$+. By time averaging the output, a precision of
0.00001 g/cm$^3$ or better could be achieved. The temperature of
the vibrating tube in the density meter was controlled with water
from a large well-regulated water bath whose temperature was
25.00$\pm$ 0.01 $^o$ C.
 \\ \\
{\bf Free-Diffusion  Measurements}. For binary Na$_2$SO$_4$-water
and ternary Lys-Na$_2$SO$_4$-Water we performed measurements for
free-diffusion using the high-precision Gosting diffusiometer
\cite{15,16,17} operated in its Rayleigh interferometric optical
mode. The procedure for measuring binary (D$_2)_v$ and ternary
diffusion coefficients (D$_{ij})_v$ were described in detail in
the work \cite{6}. In order to measure the four diffusion
coefficients of the system, the experiments must be performed with
at least two different concentration differences at each
combination of mean concentration \cite{15,18,19}. For ternary
experiments, for each pair of mean concentrations, two
measurements were performed with $\alpha_1=0$ and the two with
$\alpha_1=0.8$ ($\alpha_i$ - the
refractive index fraction due to the $i$th solute \cite{6}). \\
\noindent In order to make the data analysis of the free-diffusion
experiments we used the Fick's second law:
\begin{eqnarray}
\frac{\partial C_i}{\partial t}=\sum_{j=1}^2
D_{ij}\frac{\partial^2C_j}{\partial x^2}\;\;\;\;\;\;\;i=1,2
\end{eqnarray}
for two solutes. We made the assumption that the concentration
differences of the solutes across the initial boundary are small
enough and the diffusion coefficients are constant \cite{20}. Also
the volume changes on mixing were negligible, thus all the
measured diffusion coefficients are given relative to the
volume-fixed frame of reference defined by equation (4). \\

\noindent {\bf Dynamic Light-Scattering Diffusion Coefficients}.
We measured the dynamic light-scattering diffusion coefficients
D$_{DLS}$ for samples from all ter\-na\-ry experiments and for
0.1, 0.25, 0.5,
0.65 and 0.8 M mean concentrations of Na$_2$SO$_4$. \\
\noindent Measurements were made using a Protein Solution
DynaPro-801 TC molecular sizing instrument with a fixed scattering
angle of 90$^o$ and the procedure was described in the work
\cite{6}. This apparatus allowed us to calculate also the
eigenvalues $\lambda_1$ and $\lambda_2$ of the matrix of diffusion
coefficients, the D$_{DLS}$ predicted by Leaist's theory \cite{21}
and to compare with ours and also with interferometric value
(D$_{11})_v$. \\
\noindent {\bf The viscosity} for ternary lys-Na$_2$SO$_4$-water
system was also measured using an Ostwald viscosimeter. \\ \\ \\

\noindent {\bf Results} \\

\noindent {\bf Ternary diffusion experiments} were performed on
the lysozyme chloride-Na$_2$SO$_4$-water system at pH$=4.5$ and
T=$25^0$ C. To obtain the four ternary diffusion coefficients we
performed four experiments at the same mean concentrations but
with different values of $\Delta C_i$ for the solutes. There were
two experiments with $\Delta C_1=0$ and $\Delta C_2\neq 0$ and two
with $\Delta C_1\neq 0$ and $\Delta C_2=0$ at each mean
Na$_2$SO$_4$ concentration of 0.1, 0.25, 0.5, 0.65 and 0.8 M. The
interferometric data for the diffusion coefficients (D$_{11})_v$,
(D$_{22})_v$, (D$_{12})_v$ and (D$_{21})_v$ are reported in the
Tables 1,2,3. \\
\noindent {\bf Partial molar volumes} values, $\bar{V}_1$,
$\bar{V}_2$ and $\bar{V}_o$, were calculated for each component
using eqs A-7 ($q=2$) and 5 in \cite{22} and reported in the Tables 1,2,3. \\
\noindent Values of mean density $\bar{d}$ and $H_i=(\partial
d/\partial C_i)_{T,p,C_j,j\neq i}$ in the Table 1 were calculated
using densities of all eight solutions from each experiment set.
Densities were assumed to be linear in solute concentrations
respecting the equation \cite{6}:
\begin{eqnarray}
d=\bar d+H_1(C_1-\bar{\bar{C}_1})+H_2(C_2-\bar{\bar{C}_2})
\end{eqnarray}
where $\bar{\bar{C}_1}$ and $\bar{\bar{C}_2}$ are the averages of
the mean concentrations for all four experiments in a series. \\
\noindent {\bf Dynamic light-scattering experiments} provided the
diffusion coefficients D$_{DLS}$ for each Na$_2$SO$_4$ mean
concentrations and also the eigenvalues $\lambda_i$ ($i=1,2)$. We
could make an direct comparison to the interferometric values
(D$_{11})_v$, the smallest eigenvalues of the matrix of diffusion
coefficients $\lambda_1$ which are reported in the Table 1,2,3 and
the $D_{DLS}$ data. The values for D$_{DLS}$ are approximately 4\%
higher than (D$_{11})_v$, and approximately 5\% higher than the
smallest eigenvalue $\lambda_1$ of the matrix of
diffusion coefficients. \\
\noindent {\bf The viscosity values} for ternary solution are
reported also in the Tables 1,2,3 at each mean value of salt
concentrations. \\
\noindent {\bf The diffusion coefficients dependence} on C$_2$
mean values are shown in the Fig.1.

\begin{figure}
\includegraphics[scale=1]{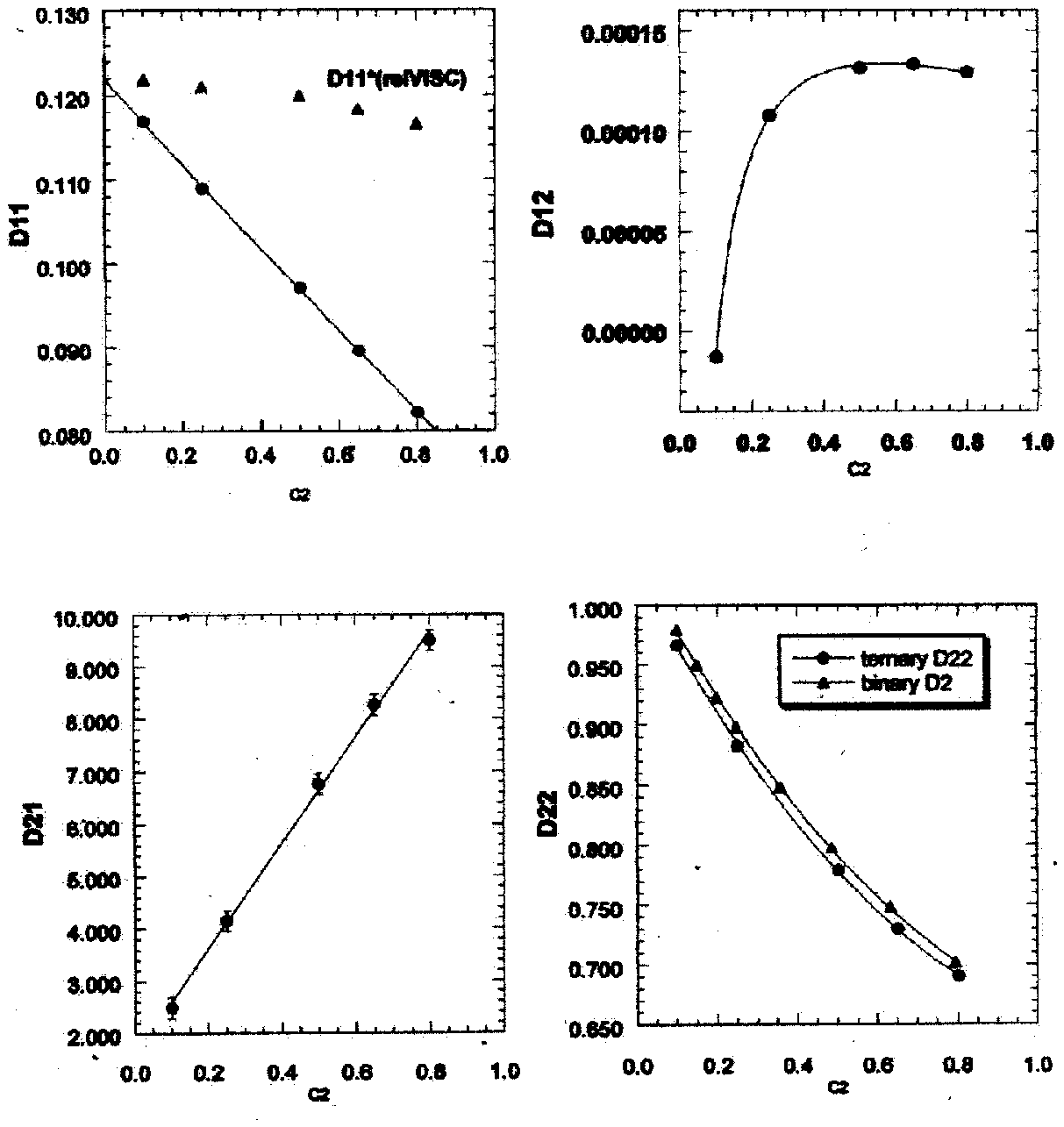}
\begin{center}
{\bf Fig. 1}
\end{center}
\end{figure}

\noindent From Fig. 1 we could see also the coefficient
(D$_{11})_v$, due to the gradient of lysozyme, corrected from
viscosity. The values for main-term (D$_{11})_v$ are 13\% smaller
than the values for the same coefficient but using the
lys-NaCl-water system, at the same concentration \cite{6}. The
cross-term (D$_{21})_v$ for the flux of Na$_2$SO$_4$ caused by the
gradient of lysozyme chloride increases as the Na$_2$SO$_4$
concentration increases. The values for (D$_{21})_v$ are 25\%
smaller than the values for the same coefficient using the NaCl as
salt \cite{6}. At 0.8 M Na$_2$SO$_4$, this term becomes 14 times
larger than the Na$_2$SO$_4$ main-term diffusion coefficient
(D$_{22})_v$. The cross-term (D$_{12})_v$ for the flux of lysozyme
caused by the gradient of Na$_2$SO$_4$ is small in comparison with
all the other diffusion coefficients as it was expected, and 35\%
smaller than (D$_{12})_v$ in the case of lys-NaCl-water \cite{6}.
Fig. 1 shows us also that the ternary main-term (D$_{22})_v$, for
the flux of Na$_2$SO$_4$ caused by the own gradient of
concentration, is smaller with 1.5\% than the binary coefficient
(D$_2)_v$, over the entire composition range, and 40\% smaller
than the (D$_{22})_v$ for ternary lys-NaCl-water \cite{6}, for the
same
range of concentration. \\ \\
\noindent {\bf Conclusions} We reported the complete set of
multicomponent diffusion coefficients for ternary
lys-Na$_2$SO$_4$-water system at concentrations high enough to be
relevant to crystallization studies. We also made a comparison
with the reported diffusion data about lys-NaCl-water \cite{6} in
order to understand the influences of the type of salt in the
diffusion process and, implicitly in the crystallization process.
\\ \\
\noindent {\bf Acknowledgment.} One of the authors (DB) is very
grateful to O. Annunziata for constant and helpful advice during
the experimental work. This research was supported by the Texas
Christian University Grant RCAF-11950 and by the NASA Microgravity
Biotechnology Program through the Grant NAG8-1356.

\vspace{1cm}
\begin{table}[h]
\begin{center}
{\bf Table 1} \\ \vspace{0.3cm}
\begin{tabular}{lcc}
$\bar{\bar{C}_1}$(mM)  & 0.6000 & 0.6000
\\
$\bar{\bar{C}_2}$(M) & 0.1000 & 0.2500  \\
$\bar{d}(g cm^{-3})$ & 1.012189 & 1.030671  \\
$H_1(10^3 g mol^{-1}$ & 4.286 & 1.149  \\
$H_2(10^3 g mol^{-1}$ & 0.12500 & 0.12230  \\
$\bar{V}_1(cm^3 mol^{-1})$ & 10050 & 10182 \\
$\bar{V}_2(cm^3 mol^{-1}) $& 17.090 & 19.780 \\
$\bar{V}_o(cm^3 mol^{-1})$ & 18.067 & 18.058 \\
$D_{DLS}(10^{-9}m^2s^{-1})$ & 0.11970 & 0.11230 \\
$\lambda_1(10^{-9}m^2s^{-1})$ & 0.11700 & 0.10840  \\
$(D_{11})_v(10^{-9}m^2s^{-1})$ & 0.1169$\pm$ 0.0001 & 0.1090$\pm$
0.0001 \\
$(D_{12})_v(10^{-9}m^2s^{-1})$ & -0.000013$\pm$ 0.000001 &
0.000108$\pm$ 0.000001  \\
$(D_{21})_v(10^{-9}m^2s^{-1})$& 2.49$\pm$ 0.01 & 4.14$\pm$ 0.01 \\
$(D_{22})_v(10^{-9}m^2s^{-1})$ & 0.9661$\pm$ 0.0001 & 0.8826$\pm$
0.0001 \\
$\eta(cp)$ & 1.042395 & 1.110034
\end{tabular}
\end{center}
\end{table}

\begin{table}[h]
\begin{center}
{\bf Table 2} \\ \vspace{0.3cm}
\begin{tabular}{lcc}
$\bar{\bar{C}_1}$(mM)  & 0.6000 & 0.6000
\\
$\bar{\bar{C}_2}$(M) & 0.5000 & 0.6500  \\
$\bar{d}(g cm^{-3})$ &  1.060538 & 1.078032
 \\
$H_1(10^3 g mol^{-1}$ & 4.104 & 4.049  \\
$H_2(10^3 g mol^{-1}$ &  0.11733 & 0.11610
 \\
$\bar{V}_1(cm^3 mol^{-1})$ & 10209 & 10257 \\
$\bar{V}_2(cm^3 mol^{-1}) $& 24.720 & 25.930
 \\
$\bar{V}_o(cm^3 mol^{-1})$ &  18.026 & 18.013
\\
$D_{DLS}(10^{-9}m^2s^{-1})$ & 0.10102 & 0.09430
 \\
$\lambda_1(10^{-9}m^2s^{-1})$  & 0.09566 & 0.08775
 \\
$(D_{11})_v(10^{-9}m^2s^{-1})$
& 0.0969$\pm$ 0.0001 & 0.0894$\pm$ 0.0001 \\
$(D_{12})_v(10^{-9}m^2s^{-1})$ & 0.000132$\pm$ 0.000001 &
0.000134$\pm$ 0.000001 \\
$(D_{21})_v(10^{-9}m^2s^{-1})$ & 6.75$\pm$
0.01 & 8.26 $\pm$ 0.01  \\
$(D_{22})_v(10^{-9}m^2s^{-1})$ & 0.7791 $\pm$ 0.0001 & 0.7294 $\pm$ 0.0001 \\
$\eta(cp)$ & 1.236427 & 1.322355
\end{tabular}
\end{center}
\end{table}

\begin{table}[h]
\begin{center}
{\bf Table 3} \\ \vspace{0.3cm}
\begin{tabular}{lc}
$\bar{\bar{C}_1}$(mM)  & 0.6000
\\
$\bar{\bar{C}_2}$(M) &  0.8000 \\
$\bar{d}(g cm^{-3})$ &  1.095745 \\
$H_1(10^3 g mol^{-1}$ &  3.954 \\
$H_2(10^3 g mol^{-1}$ & 0.11502 \\
$\bar{V}_1(cm^3 mol^{-1})$ &  10341\\
$\bar{V}_2(cm^3 mol^{-1}) $& 26.980 \\
$\bar{V}_o(cm^3 mol^{-1})$ & 17.993 \\
$D_{DLS}(10^{-9}m^2s^{-1})$ & 0.08683 \\
$\lambda_1(10^{-9}m^2s^{-1})$ &  0.08021 \\
$(D_{11})_v(10^{-9}m^2s^{-1})$ & 0.0822$\pm$ 0.0001 \\
$(D_{12})_v(10^{-9}m^2s^{-1})$ & 0.000130$\pm$ 0.000001 \\
$(D_{21})_v(10^{-9}m^2s^{-1})$& 9.50$\pm$ 0.01 \\
$(D_{22})_v(10^{-9}m^2s^{-1})$ & 0.6900$\pm$ 0.0001 \\
$\eta(cp)$ &  1.418206
\end{tabular}
\end{center}
\end{table}

\end{document}